**Asymptotic emergence of statistically supported false causal interpretation under unmeasured confounding**

Mark Louie F. Ramos, Ph.D.

Department of Health Policy and Administration, The Pennsylvania State University, University Park, PA mlr6219@psu.edu

Abstract

Unmeasured confounding is widely recognized as a limitation of observational causal inference, but its implications for data-driven causal discovery are often understated. We provide an asymptotic characterization of this limitation. When the true causal variable is unobserved but correlated proxy variables are available, increasing sample size while searching for increasingly stable, data-driven causal structures will make associations involving observable proxies become increasingly certain. This result highlights a fundamental distinction between reducing statistical uncertainty and recovering causal mechanisms: more data can improve certainty about genuine relationships within the observed variable space while providing no guarantee of identifying the underlying causal structure.

## Introduction

As observational causal inference methods become more common across applied fields, communicating their assumptions and the limits of what they can establish is increasingly critical. Well-understood in causal inference theory is that observational causal methods do not determine whether a proposed cause–effect relationship holds. Rather, they assume a causal structure, typically formalized through tools such as directed acyclic graphs (DAGs), instrumental variables, or quasi-experimental designs, and estimate causal effects conditional on those assumptions (Hernan & Robins, 2024; Pearl, 2014).

Several results stem from this principle. For example, it has been shown that different causal models can be empirically indistinguishable (Verma & Pearl, 1991). This means that, in practice, observational data alone cannot determine which causal structure generated the data. Connected to this, it has been discussed that statistical tests in observational causal analyses evaluate properties of an assumed causal model, and are not meant to test the validity of the causal structure itself (Greenland et al., 1999). Another well-known result is the issue of unmeasured confounding, such that if there exist unobserved common causes of treatment and outcome, then causal effects cannot, in general, be recovered from observational data alone (Shpitser, 2020).

Despite the wide publication of these results, problems rooted to sufficient appreciation of the importance of these assumptions are commonplace. Application of causal inference on observational settings frequently proceed without clearly stating or defending the identifying assumptions on which they rely (Hernan & Robins, 2024). A systematic review found that assumptions encoded by researchers in their DAGs about the data-generating process are often left implicit or under-reported (Tennant et al., 2021).

This paper contributes a result that shows a complementary perspective to the theoretical discussion of unmeasured confounding. We consider the setting in which the true causal relationship is fundamentally unobservable, yet the demand for causal answers motivates the collection of increasingly rich observational datasets. As observational datasets grow richer and estimation becomes more precise, workflows that map statistical association to causal interpretation will inevitably support variables with stable, statistically significant effects, even when none of those variables are causal parents of the outcome.

The purpose of this exposition is not to introduce new results on causal identification nor to discourage the use of observational causal inference methods, but to underscore the underappreciated point that observational causal inference must begin with convincing assumptions about which the purpose is measurement, not validation. The operative question is not Does X cause Y? But rather suppose X causes Y, what is the magnitude of that effect?

**Setup**

(1) Let $V$ be the set of all variables in the universe and $G_{true}$ be the true directed acyclic graph (DAG) over $V$.
(2) Let $O \subset V$ be the set of observable variables.
(3) Let there exist $X, Y \in V$ such that $X \to Y$ in $G_{\text{true}}$ with total effect $\rho$, and $X \notin O$.
(4) Let there exist a sequence $\{Z_j\}_{j=1}^{\infty} \subset O \setminus \{Y\}$ such that:

- $X \to Z_j$ in $G_{\text{true}}$ for all $j$,
- there is no edge $Z_j \to Y$ in $G_{\text{true}}$ for any $j$,
- $\text{Cov}(X, Z_j) \neq 0$ for all $j$ and $\text{Cov}(Z_j, Y) \neq 0$. (i.e. $Z_j$ is a non-causal proxy of $X$)
- Each $Z_j$ is a candidate focal variable when observed

(5) Let $\{O_k\}_{k=1}^{\infty}$ be increasing subsets of $O$ with $\bigcup_k O_k \supset \{Z_1, Z_2, \dots\}$. For each $k$, let $n$ be the sample size of i.i.d. data from the marginal distribution over $(Y, O_k)$.
(6) For each $k, n$, a procedure $\mathcal{A}_{k,n}$ takes the data on $(Y, O_k)$ and, for any focal variables $W_{\text{i},k} \in O_k$ it tests, using:

- a consistent estimator of $\theta_{\text{i}}$, the association between $W_{\text{i},k}$ and $Y$, which is interpreted as a causal effect under the assumed DAG.
- a standard test of $H_0: \theta_{\text{i}} = 0$ vs $H_1: \theta_{\text{i}} \neq 0$,
- If the test rejects, it is interpreted as "$W_{\text{i},k}$ (at least partly) causes $Y$" under the assumption of no unmeasured confounding between $W$ and $Y$.

**Claim**

For a sufficiently large $k$, $\Pr\left(\mathcal{A}_{k,n} \text{ concludes } Z_j \to Y\right) \underset{n\to\infty}{\to} 1.$

**Proof**

Statement 1:

$$\Pr(W_k \in \{Z_1, Z_2, \dots\}) \underset{k\to\infty}{\to} 1$$

Reason: By construction, $\{O_k\}$ in $\mathcal{A}_{k,n}$ is increasing and $\bigcup_k O_k$ contains all $Z_j$, so for large $k$, at least one $Z_j$ is in $O_k$. Since every $Z_j$ is a candidate focal variable, once they are in $O_k$, they are included as one of the hypothesis tests for $\mathcal{A}_{k,n}$.

Statement 2: For $Z_j$ of interest in $\mathcal{A}_{k,n}$ , $\mathrm{Cov}(Z_j, Y) = \theta_i \neq 0$

Reason: Given (4)

Statement 3: $\Pr\left(\mathcal{A}_{k,n} \text{ rejects } H_0\colon \theta_i = 0\right) \underset{n\to\infty}{\to} 1.$

Reason: Definition of statistical power and consistent estimators

Statement 4: For a sufficiently large $k$ (such that $Z_j \in O_k$ for some $j$), $\Pr\left(\mathcal{A}_{k,n} \text{ concludes } Z_j \to Y\right) \underset{n\to\infty}{\to} 1.$

Reason: Follows from Statement 3 and Given (6)

**Discussion**

Except for the characteristic that $X \notin O$, the setup mirrors common, good-faith practice in scientific research with observational data. Researchers typically begin with a substantive outcome of interest $Y$ and a hypothesized causal factor, informed by theory, prior literature, or rational plausibility. Just because the true cause $X$ is unobservable, even unknown, does not mean there is a dearth of $Z_j$'s that are, themselves, plausible causes of $Y$ even if none of them actually are. This is not new and is in fact the thrust of results around unmeasured confounding. What my result makes explicit is that in this scenario, statistical principles will continue to apply on the estimators regardless if their estimands are truly causal, partial causal, or non-causal effects. Moreover, as datasets become "better" in terms of both how many features we are able to capture about the data generating process ($i.e. \{O_k\}_{k\to\infty}$) and how precise we are at estimating estimands given those features (i.e. $\{\mathcal{A}_{k,n}\}_{n\to\infty}$), we will inevitably find statistically sound evidence of association that will incorrectly be concluded as causal association. This result is not meant to discourage the use of observational causal inference methods, it is only emphasizing that there are two, equally important parts to it and statistical analysis of observational data only plays a role in the latter part. The first part, declaring assumptions about the underlying causal structure, is entirely informed by prior knowledge, which may include insights from previous results of statistical inference, but not validated by subsequent statistical inference on the present study's data. Rather, statistical inference is made conditioned on those assumptions being true. Thus, regardless of the results that follow from estimation and testing, those assumptions remain as reasonable (or unreasonable) as they were prior to estimation. What statistical analysis provides is that stakeholders now have an estimate of what the causal effect might be suppose the assumptions made are true.

It is important to point out that this is not a statistical issue. That is, it does not stem from poor practice of Type 1 error rate control like in issues of p-hacking, data dredging, cherry-picking, or failure to account for multiple comparisons which are known to negatively impact the quality of statistical inference (Wasserstein & Lazar, 2016). In fact, since $\mathrm{Cov}(X, Z_j) \neq 0$ and $\mathrm{Cov}(Z_j, Y) \neq 0$ for all $j$, replicating $\{\mathcal{A}_{k,n}\}$ that rejected its $H_0\colon \theta_\mathrm{i} = 0$ at the same or higher sample size is likely to result in the same outcome. That is, the results are more likely to be replicated than not, since the association is actually real, it's just not causal.

It is likewise critical to distinguish between assumptions necessary to do the first part of observational causal inference versus statistical assumptions made when doing the latter part. Whereas neither class of assumptions are exactly verifiable, statistical assumptions tend to be more robust, such that the impact of violations on inference depends on the gravity of the violations which are, to some extent, perceptible from the data. As such, so-called "validation" procedures are reasonably sufficient to rule out egregious issues. For example, the basic t-test requires the data-generating process (dgp) to be exactly normally distributed to achieve exact, level-alpha control over Type I error. In reality, no dgp is ever exactly normal, and normality-checking procedures are at best crude diagnostics. Yet the t-test itself is highly robust to violations of normality, such that moderate skew or kurtosis typically has only minor effects on inference, especially with adequate sample sizes. In contrast, the assumptions of a correctly specified DAG and no unmeasured confounding are rigid. While some violations can have minimal impact. For example, an unmeasured confounder may only weakly affect both exposure and outcome such that the estimated effect is just biased in magnitude, it is also possible for a hidden cause to entirely drive both, making it incorrect to attribute any effect as causally contributed by the exposure. Observational data cannot distinguish between these situations. Sensitivity analyses help in quantifying how large a violation would need to be to overturn the estimated effect, but they still operate under the assumption that no unmeasured confounder is wholly responsible for the observed association instead of the presumed causal variable.

Naturally, assuming $X \notin O$ is overly pessimistic. After all, the point of observational causal inference is ultimately to provide recommendation on whether or not it is reasonable to act as though implementing changes in $X$ causes changes in $Y$. It was done to emphasize the point that statistically convincing estimation of assumed causal estimands is not just possible, but asymptotically likely under increasingly better data regardless of whether the estimand is truly causal.

Ultimately, these results provide an asymptotic characterization of why the standard caveat about unmeasured confounding has a stronger consequence than is often appreciated. When the true causal variable is unobserved but correlated proxies are available, increasing sample size does not resolve the causal ambiguity. Instead, it increases statistical confidence in relationships involving the observed proxies. Consequently, strategies that seek to discover increasingly stable, data-driven DAGs through larger observational datasets can converge toward causal explanations that accurately capture associations in the observed data while incorrectly attributing causality to proxy variables rather than to the unobserved mechanism. This invites researchers, reviewers, and readers to regard observational causal assumptions with at least as much care and scrutiny as the data analysis itself.